\documentclass[sigconf, 9pt]{acmart}

\usepackage[english]{babel}
\usepackage{graphicx} 
\graphicspath{{figures/}} 

\renewcommand\footnotetextcopyrightpermission[1]{} 
\setcopyright{none}

\acmYear{2021}\copyrightyear{2021}
\setcopyright{acmcopyright}
\acmConference[NAI '21]{ACM SIGCOMM 2021 Workshop on Network-Application Integration}{August 27, 2021}{Virtual Event, USA}
\acmBooktitle{ACM SIGCOMM 2021 Workshop on Network-Application Integration (NAI '21), August 27, 2021, Virtual Event, USA}
\acmPrice{15.00}
\acmDOI{10.1145/3472727.3472797}
\acmISBN{978-1-4503-8633-3/21/08}

\begin{document}
\title{Wide Area Network Autoscaling for Cloud Applications}

\author{Berta Serracanta} 
\authornote{Both authors contributed equally to this work.}
\email{bserraca@cisco.com}
\affiliation{%
  \institution{UPC-BarcelonaTech}  \city{Barcelona}  \country{Spain}
}\affiliation{%
  \institution{Cisco}
  \city{San Jose}
  \country{USA}
}

\author{Jordi Paillisse}
\authornotemark[1]
\email{jordip@ac.upc.edu}
\author{Albert Cabellos}
\email{acabello@ac.upc.edu}
\affiliation{%
  \institution{UPC BarcelonaTech}
  \city{Barcelona}
  \country{Spain}
  }


\author{Anna Claiborne}
\email{aclaiborne@packetfabric.com}
\affiliation{%
 \institution{PacketFabric}
 \city{Los Angeles}
 \country{USA}
 }

\author{Alberto Rodriguez-Natal}
\email{natal@cisco.com}
\affiliation{%
 \institution{Cisco}
 \city{San Jose}
 \country{USA}
 }
 
\author{Dave Ward}
\email{dward@packetfabric.com}
\affiliation{%
 \institution{PacketFabric}
 \city{Los Angeles}
 \country{USA}
 }

\author{Fabio Maino}
\email{fmaino@cisco.com}
\affiliation{%
  \institution{Cisco}
  \city{San Jose}
  \country{USA}
  }

\renewcommand{\shortauthors}{Serracanta et al.}

\begin{abstract}

Modern cloud orchestrators like Kubernetes provide a versatile and robust way to host applications at scale. One of their key features is autoscaling, which automatically adjusts cloud resources (compute, memory, storage) in order to adapt to the demands of applications. However, the scope of cloud autoscaling is limited to the datacenter hosting the cloud and it doesn't apply uniformly to the allocation of network resources. In I/O-constrained or data-in-motion use cases this can lead to severe performance degradation for the application. For example, when the load on a cloud service increases and the Wide Area Network (WAN) connecting the datacenter to the Internet becomes saturated, the application flows experience an increase in delay and loss. In many cases this is dealt with overprovisioning network capacity, which introduces  additional costs and inefficiencies.

On the other hand, thanks to the concept of "Network as Code", the WAN  exposes a  set of APIs  that can be used to dynamically allocate and de-allocate capacity on-demand. In this paper we propose extending the concept of cloud autoscaling into the network to address this limitation. This way, applications running in the cloud can communicate their networking requirements, like bandwidth or traffic profile, to a Software-Defined Networking (SDN) controller or Network as a Service (NaaS) platform.  Moreover, we aim to define the concepts of vertical and horizontal autoscaling applied to networking. We present a prototype that automatically allocates bandwidth  to the underlay network, according to  the requirements of the applications hosted in Kubernetes.   Finally, we discuss open research challenges.

\end{abstract}



\maketitle

\section{Introduction}

During the last decade, Software-Defined Networking (SDN) has enabled programmability to network management. By exposing expressive application-facing APIs, the underlying  network  can be controlled dynamically, offering  quasi real time network control. Recently, several commercial networking scenarios have embraced the SDN approach, resulting in a new breed of networking solutions, such as SD-Wide Area Network \cite{B4paper} or SD-Access Networks \cite{sdaPaper}.

Starting even earlier and driving the as-a-Service revolution, the \emph{compute} environment has been following its own path of  innovation. Compute resources have become more granular, from physical hardware to Virtual Machines (VMs), then to containers \cite{containers} and more recently towards Lambda functions. This  allows applications to  consume compute resources more efficiently, as well as the level of scalability that has made Cloud Computing possible.


In this context, orchestration software to manage large pools of compute resources, such as OpenStack for VMs \cite{openStackArchitecture}, or Kubernetes for containers \cite{containers} has become commonplace. These orchestration infrastructures have been naturally extended into the networking domain, especially in the case of service meshes, such as Envoy proxies controlled by  Istio \cite{serviceMeshSoA, PacketFabricNaasBlog}. However,  the capability of the network to keep up with the scalability  offered by the cloud infrastructure is limited mostly to datacenters (within the cluster). Inter-cluster connectivity, as well as the connection with the users, is typically provided via Wide-Area Networks (WAN), a relatively expensive resource with limited capacity.

In addition to the limited capacity of the WAN, the lack of communication between the cloud and the WAN hinders some of the advantages of the cloud.  A key limitation  emerges  in the context of cloud autoscaling events. Autoscaling is a cloud feature to dynamically adapt the amount of compute and memory resources to the current application load, as a way to improve efficiency and optimize cost.


In current cloud deployments that span across  datacenters and WANs, autoscaling events detected by the cloud orchestrator only propagate within the compute domain (i.e., datacenter). Although an application can scale up to support higher loads, the connection towards the users or between clusters remains unchanged, which might cause congestion or reduce Quality of Experience (QoE). It may also require overprovisioning of network capacity, introducing significant additional costs and inefficiencies.  Conversely, when applications scale down, network resources become idle and underutilized. Taking this into account,  we argue that there is a need to develop  abstractions and protocols to interface the \emph{compute} and the \emph{networking} domains, with a strong focus on the network autoscaling properties.

In this paper we explore how application-driven cloud autoscaling events can be proactively  propagated  to the network. Common approaches to network autoscaling are reactive, i.e., they monitor network traffic, compute an average, and trigger autoscaling events if the average exceeds a threshold \cite{trafficPredictionARMA}. However, in our case we make use of application  context extracted from the cloud to proactively autoscale the network. This approach doesn't wait for surges in  network load to increase capacity, but rather proactively increases network capacity shortly before surges in demand become observable in the network. With WAN autoscaling, the network dynamically matches the requirements of applications with the resources available in the network, achieving higher efficiency in resource utilization.

The rest of this paper is organized as follows. First, we define vertical and horizontal network autoscaling. Then, we present two use cases of vertical and horizontal network autoscaling, and experimentally evaluate their performance. Finally, we derive open challenges for future research in the field.



\section{Network Autoscaling}

Autoscaling is a cloud feature to make efficient use of compute and memory resources by automatically adapting their allocation  to the current application load. This is done either by adding/removing more resources to an application instance (vertical autoscaling) or by adding/removing new application instances (horizontal autoscaling).



\begin{figure}[!tb]
\includegraphics[width=\columnwidth]{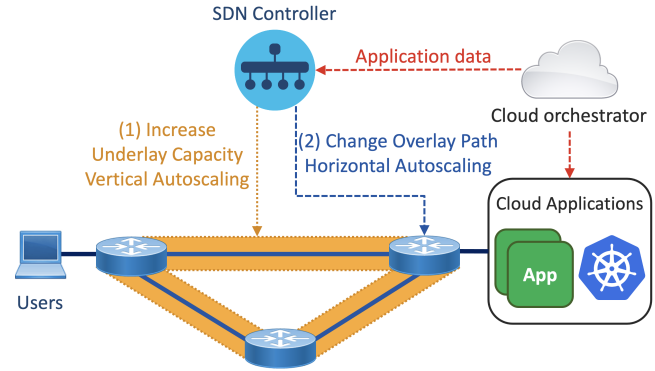}
\centering
\caption{Vertical and Horizontal Network Autoscaling}
\label{fig:as_concepts}
\vspace{-0.5cm}
\end{figure}

Therefore, we propose extending these two concepts to the networking domain in the following way (fig. \ref{fig:as_concepts}): first, we make use of the APIs exposed by   cloud orchestrators (e.g., Kubernetes) to extract network requirements from the applications running in the cloud. In cloud environments, it is common that developers add labels or annotations to workloads in order to ease management and automate tasks \cite{kubernetesAnnotations}. We take advantage of this widely-used feature so developers can add information about the network requirements of the applications. In turn, we extract application-level context available in the cloud. For example, we can detect cloud autoscaling events that indicate that the demand for a given application is increasing and may require more network capacity. 

Then, we send this information to an SD-WAN controller so it can be programmed by NetOps to act upon the network. For example (fig. \ref{fig:as_concepts}), we can request more bandwidth  over a specific connection  to an underlay provider (1), or change some paths according to the requested traffic profile (2). 

Note that we aim to clearly separate the jobs of DevOps and NetOps with a simple interface. On one hand, DevOps make use of cloud tools to add context information about the application. In our case, they add annotations to define the traffic profile needed by a given application (annotations are typically fairly abstract indications about the application network behavior using labels such as “file transfer”, “video streaming”, or ”database analytics”). On the other hand, NetOps can automate the optimal allocation of network resources to applications thanks to their in-depth knowledge of the WAN infrastructure. In the next sections, we describe in detail the concepts of vertical and horizontal network autoscaling.

\subsection{Vertical Network Autoscaling}
In the cloud, vertical autoscaling modifies the properties of an existing compute instance. Similarly, we define Vertical Network Autoscaling as changing the properties of an existing connection (e.g., Layer 2 pseudo-wires, MPLS tunnels, VXLAN (RFC 7348) or LISP tunnels (RFC 6830), etc). For example, if the number of instances of a specific application increases, we will increase accordingly the bandwidth of the connection serving these applications.

We are assuming that: (i) the resource limiting application performance is the amount of WAN bandwidth, and (ii) the SD-WAN controller has access to an API-driven underlay provider that can dynamically provision the capacity of the connections, as it is becoming common with NaaS providers \cite{PacketFabricNaasBlog}.


\subsection{Horizontal Network Autoscaling}
Taking into account that cloud horizontal autoscaling means adding more instances to handle growing demand, we define Horizontal Network Autoscaling as changing the path inside the network that a specific application is currently following, due to changes in its requirements.  For instance, consider that a video streaming application can tolerate latency up to a certain maximum, and that DevOps have labeled such application in the cloud orchestrator. We can use these labels to pull information from the orchestrator to identify this application in the network,  e.g., via IP and port. Since most SD-WAN controllers  monitor  path properties like delay, jitter or bandwidth, we can use this information to steer application flows through the appropriate paths in the SDN network  via traffic engineering or Segment Routing (SR).

However, note that when we change the path inside the network we do not necessarily mean inside the \emph{same} network. In other words, we can dynamically choose from different providers to adapt better to sudden increases in load, using resources more efficiently. This is especially relevant from a business perspective, because increasing efficiency translates to reductions in cost through price arbitrage. For example, consider a virtual circuit of 1 Gbps that can be extended up to 2 Gbps, but the price of this additional Gbps is significantly higher than the first. Upon a surge in application traffic, instead of contracting this additional Gbps that is more expensive, we can request it to to a different provider that offers a less expensive alternative. More details about this can be found in Section \ref{Conclusions}.

\section{Use Case I: Vertical Network Autoscaling}
\label{sec:useCaseVertical}

\subsection{Scenario}

\begin{figure}[!tb]
\includegraphics[width=\columnwidth]{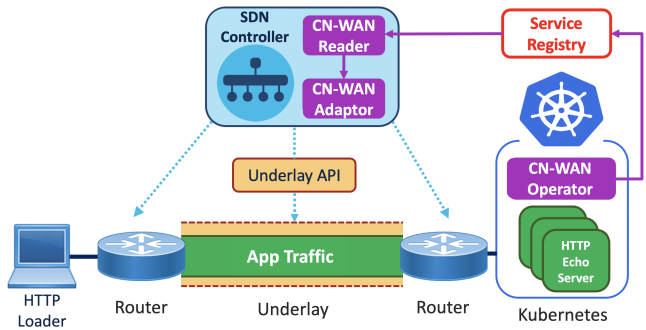}
\centering
\caption{Scenario overview}
\label{fig:Overview}
\end{figure}

The experimental setup consisted of three main parts:  cloud, network, and cloud-network API (fig. \ref{fig:Overview}).

\subsubsection{Cloud}
We used a public cloud for our experiments \cite{GCloud} controlled by Kubernetes (Version 1.18.16-gke.502), and an ad-hoc HTTP Echo server \cite{echo-server} as a cloud application. 


We leveraged the functionalities of the Kubernetes Horizontal Pod Autoscaler (HPA) to generate autoscaling events. The HPA is based on a simple algorithm that operates between a target metric and the current value of the metric, adding or deleting replicas if the current state does not match the desired state.
In our implementation, the target metric was the average CPU utilization across all monitored pods. We set the HPA threshold to 40\%, meaning that when CPU utilization is greater than 40\%, Kubernetes deploys new containers, scaling from 1 up to 150 replicas.
We used a custom HTTP Loader to increase the load on the HTTP Echo server in the Kubernetes containers. The HTTP Loader artificially generated 700 HTTP echo requests, and we increased this number to 1900, 2900, 3500, and 4500  connections, at 30, 90, 150, 210 seconds since the start of the test, respectively.





\subsubsection{Network}
We used a commercial underlay network connecting Washington D.C. and Seattle based on a Segment Routing tunnel (RFC 8402), with a baseline of 50 Mbps and a maximum capacity of 100 Gbps. This network is programmatically controlled by a proprietary API \cite{PacketFabricNaasBlog}. Fundamentally, the API allows to create, upgrade and delete virtual circuits between the two established locations. This network connected the client (HTTP Loader in Washington D.C.) with the cloud application (HTTP Echo server in Seattle).

\subsubsection{Cloud-Network API}
We developed an open-source API that allows the network and the cloud to talk, the Cloud-Native SD-WAN project (CN-WAN \cite{CN-WAN}). The main role of CN-WAN is to provide an interface between cloud applications and  networks that connect to end-users, or applications running across multiple clusters. This interface allows us to: (i) identify which cloud applications require network autoscaling, (ii) communicate autoscaling events from the cloud to the network, and (iii) quantify the scaling factor required on the underlay.

The CN-WAN project consists of three main blocks: the Operator, the Reader and the Adaptor (fig. \ref{fig:Overview}). When Kubernetes detects a variation in the number of replicas of the application being monitored, the Operator publishes this change to an external service registry. From there, the Reader polls the service registry and announces the events to the Adaptor. Specifically, the Reader-Adaptor interface leverages an HTTP-based RESTful API and transmits metadata about a specific endpoint or application in the form of key-value pairs, e.g., name: ApplicationName and replicas: 5. Finally, the Adaptor contacts the programmable underlay to adjust the network to the upcoming bandwidth requirements.

The different modules of the CN-WAN work independently, that is, they can have a standardized API to send/receive data from different sources. For example, we can insert data into the Reader that does not necessarily come from Kubernetes, e.g., from a private DNS containing information from application endpoints. On the adaptor side, we can have different adaptors depending on the underlying SD-WAN controller or NaaS platform, but all the adaptors use the same RESTful API when listening to events from the Reader.

Finally, note that  we are assuming that an increase in CPU utilization translates to an increase in the required bandwidth, and that the application developer or network administrator has an accurate knowledge about the bandwidth required by each application replica. Although this can be helpful given that the developer has a deep understanding of the application, we could also combine this knowledge with more refined techniques, such as time-series based traffic prediction models \cite{time-series-traffic-prediction}. Section \ref{Conclusions} (Bandwidth Estimation) discusses how to estimate the application bandwidth.



\subsection{Evaluation}

\subsubsection{Bandwidth Autoscaling}

\begin{figure}[!tb]
\includegraphics[trim = {0.2cm 0.1cm 0.2cm 0cm}, clip,width=\columnwidth]{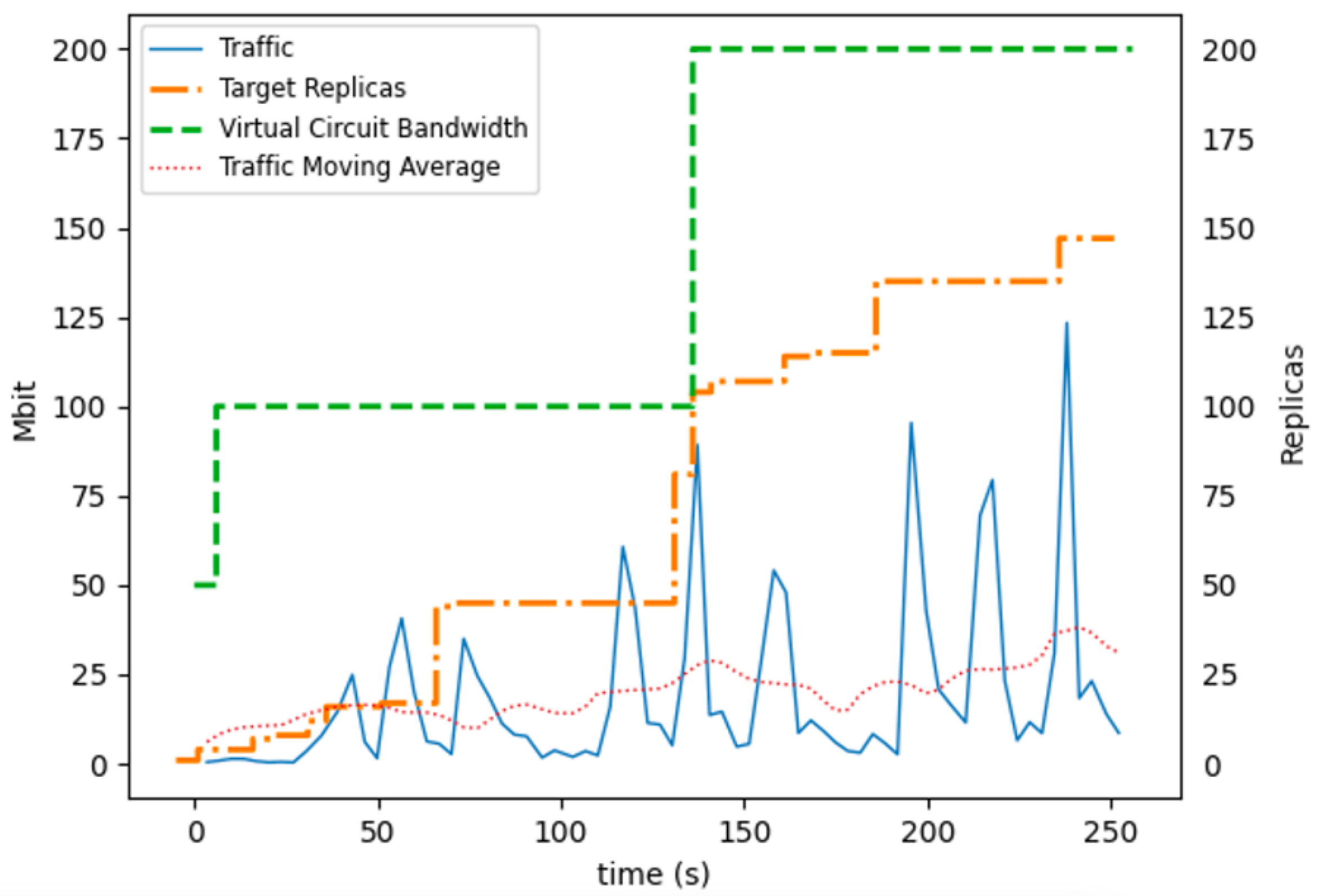}
\centering
\caption{Vertical autoscaling performance}
\label{fig:vertical}
\vspace{-0.5 cm}
\end{figure}

Fig. \ref{fig:vertical} shows an overview of the operation of vertical network autoscaling. We have measured the throughput in the cloud router connected to the virtual circuit, the total capacity of the connection, and the number of application replicas in Kubernetes. The plot shows how the bandwidth allocated in the underlay's virtual circuit starts autoscaling proactively with the application to support  upcoming traffic.


More specifically, we started injecting traffic to a 50 Mbps virtual circuit to generate autoscaling events. We can see that traffic starts increasing around 50 seconds after the experiment starts (blue line), that simulates a spike in demand for  a particular cloud application. In parallel, as  traffic increases, the CPU utilization of the cloud application rises, and Kubernetes automatically deploys more replicas (orange line, at the same time as traffic increases). We can see that shortly after the number of replicas is greater than  50, the underlay capacity is increased from 100 Mbps to 200 Mbps (at around 140 seconds in the timescale). This is because we trigger autoscaling events when traffic is greater than the minimum allocated bandwidth in the virtual circuit (50 Mbps). We estimate traffic assuming that each replica consumes a fixed amount of bandwidth,  1 Mbps in this case. Since at this moment there are aprox. 75 replicas, this translates to 75 Mbps, so we jump to the next bandwidth step allowed by the underlay provider (here from 100 to 200 Mbps).
Another example of a network autoscaling event can be seen at the beginning of the experiment. The initial jump in the virtual circuit bandwidth from 50 to 100 Mbps is caused by the combination of the minimum bandwidth (50 Mbps) of the virtual circuit and the first autoscaling event on the cloud (1 replica), which triggers a jump to first granularity step allowed by the underlay provider (50 Mbps).






We must remark that the increase in virtual circuit capacity is triggered proactively, as soon as the cloud application starts autoscaling (the sudden increase of 50 to 75 replicas is very close in time to the increase of virtual circuit bandwidth). This means that the underlay virtual circuit capacity is increased \emph{before} the actual network traffic hits the 100 Mbps initial capacity of the underlay, i.e,  the blue line never crosses the green one. 

Furthermore, as soon as the cloud application scales down because of diminished demand, the network will scale down the underlay capacity accordingly (not shown in the graph), resulting in a highly efficient and cost effective usage of the underlay resources. This example shows the benefits of autoscaling the WAN, compared with today’s typical approach of overprovisioning underlay capacity to address peaks of demand, or common autoscaling techniques that leverage traffic prediction models based on autoregression or moving average to decide when to scale up or down \cite{trafficPredictionARMA}.

\begin{figure}[!tb]
\includegraphics[trim = {0.4cm 0.6cm 1.5cm 1cm}, clip,width=\columnwidth]{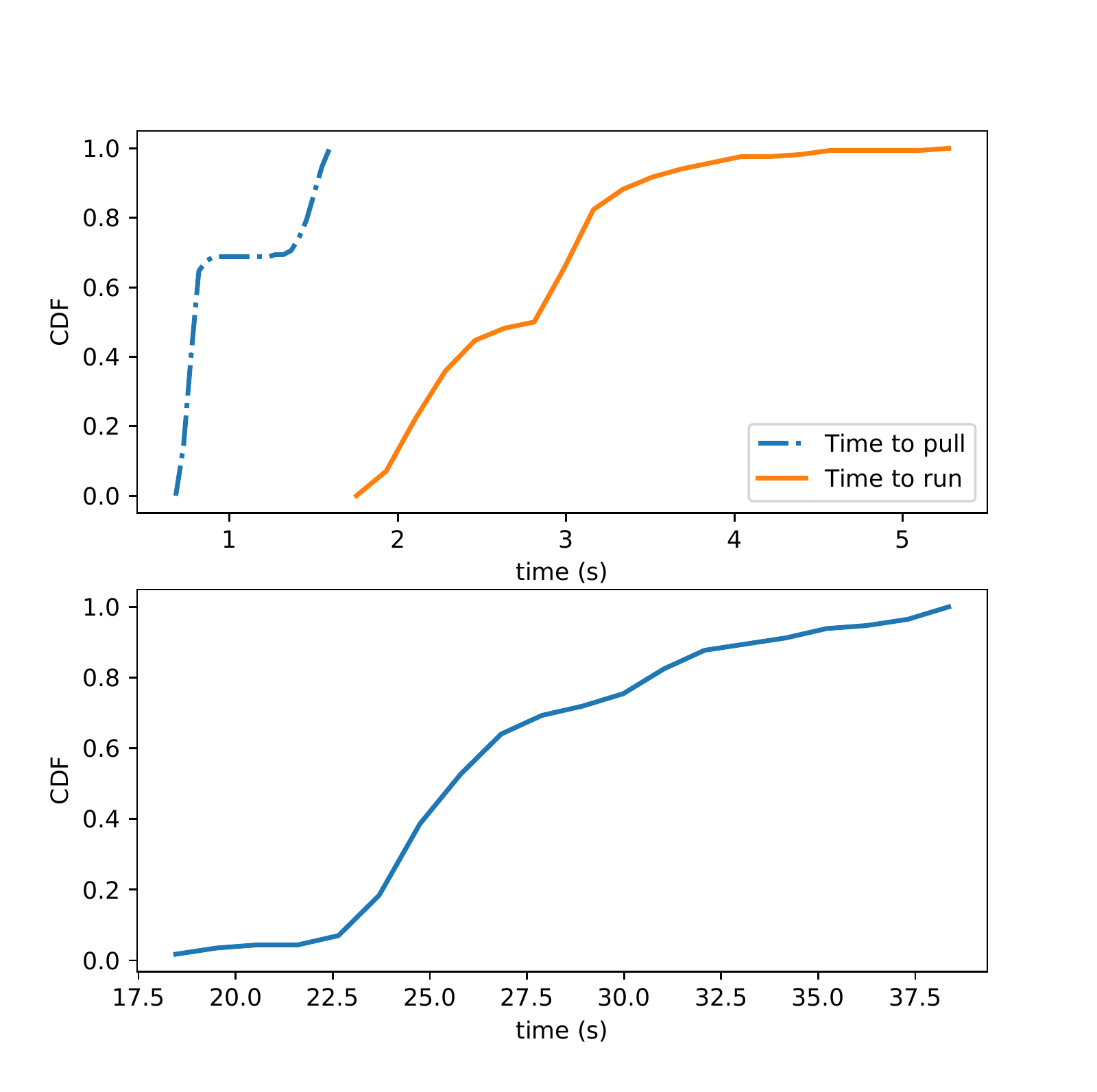}
\centering
\caption{Kubernetes controller (top) and overlay controller (bottom) response time}
\label{fig:delay}
\vspace{-0.5cm}
\end{figure}

\subsubsection{Response Time}

Next, we focus on the response time of  Kubernetes and the overlay controller when reacting to  autoscaling events. The top plot in fig. \ref{fig:delay} shows the Cumulative Distribution Function (CDF) of the Kubernetes Horizontal Pod Autoscaler (HPA) latency in two situations. \emph{Time to pull} corresponds to the delay to pull the container image when we receive a new autoscaling event. 
\emph{Time to run} is the delay between  the trigger of an autoscaling event by the HPA until Kubernetes indicates that the new container is running. Note that the latter includes the time to pull the image, and that each CDF contains 170 measurements. In both cases, we can see that Kubernetes can autoscale in the order of seconds.



Finally, the bottom plot in fig. \ref{fig:delay} shows the CDF of the delay of the SD-WAN overlay controller, i.e., the time between receiving an autoscaling event in the overlay until the capacity of the virtual circuit is updated. We can see that this time is in the order of tens of seconds. Although this delay is one order of magnitude above the delay of the Kubernetes controller, it is remarkably  lower than current  WAN circuits, a process that can take days in most of the cases. The provisioning delay is further discussed in Section \ref{Conclusions} (Provisioning Time).



\section{Use Case II: Horizontal Network Autoscaling}

We modified the setup of Section \ref{sec:useCaseVertical} in order to steer a traffic flow between two different underlay tunnels. More specifically: (i) instead of the HTTP echo server, we deployed a container that streamed video from the Kubernetes cluster, (ii) we connected a VM to receive the video stream on the other end of the WAN, (iii) we annotated the Kubernetes container with a label to identify it as a video stream, and (iv) we configured the SD-WAN controller with two tunnels: one with a limited bandwidth of 3 Mbps, an another with 1 Gbps. 

This way, the CN-WAN operator reads the IP address and port number(s) of the video stream container from Kubernetes, as well as metadata attached to the container. As we mentioned previously, this metadata is added by the application developer in the form of Kubernetes annotations. In this case, the metadata is a key-value pair indicating the traffic profile of the endpoint (\texttt{traffic-profile: video}).

The operator stores this information in a service registry, and the reader collects this information and sends it to the SD-WAN controller via the adaptor. The reader uses an API common for any type of adaptor, that indicates the IP address, port number, and traffic profile of the Kubernetes application.

Then, the adaptor configures the SD-WAN policies to route the traffic of each endpoint through a specific WAN tunnel, depending on the associated traffic profile. The mapping between traffic profile and tunnel has been defined by NetOps, in order to satisfy the specific requirements of each traffic profile in terms of bandwidth, delay, etc. For example, the traffic profile "video" maps to SD-WAN tunnels with high bandwidth, while the profile "standard" maps to a tunnel going over the public Internet. 

\begin{figure}[!ht]
\includegraphics[trim = {0.6cm 1.8cm 1.5cm 0cm}, clip,width=\columnwidth]{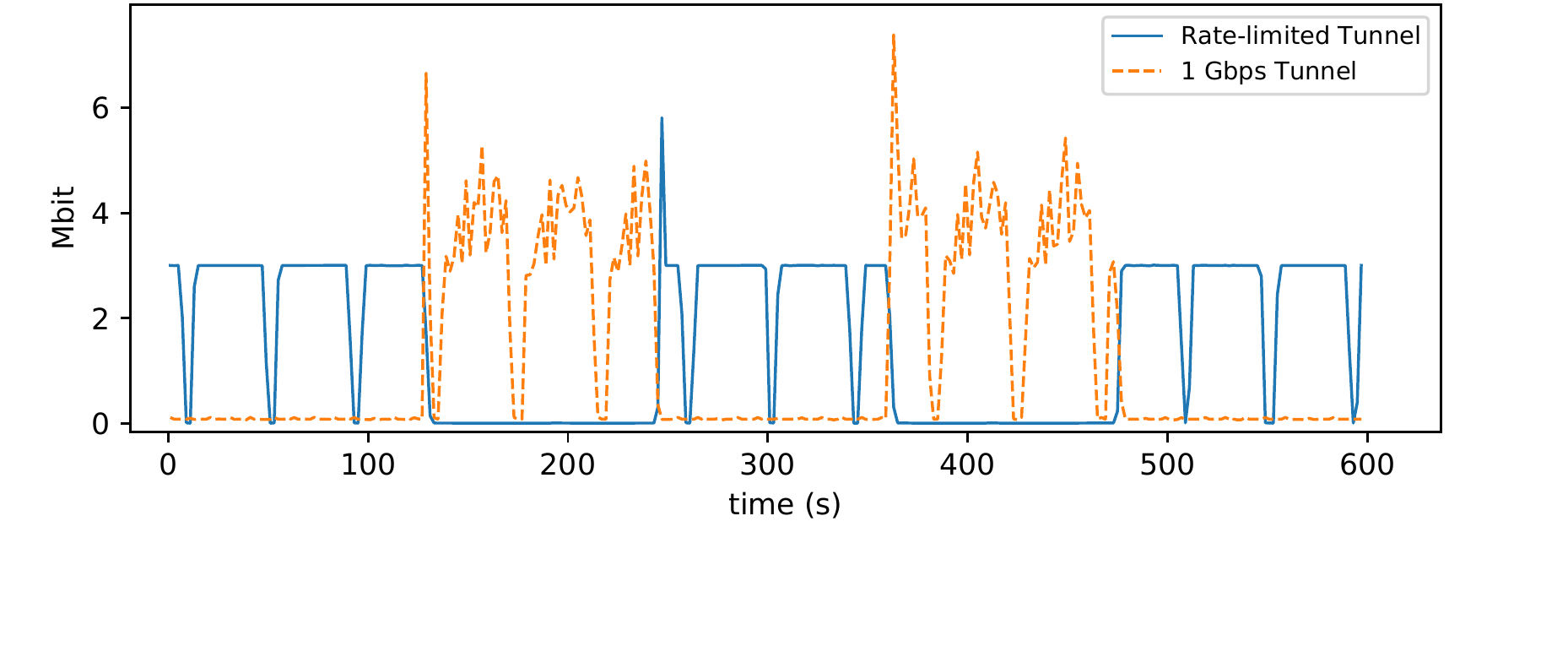}
\centering
\caption{Throughput in the rate-limited tunnel and the 1 Gbps tunnel}
\label{fig:tunnel_bw_switch}
\vspace{-0.4cm}
\end{figure}

In order to change the path of the video stream, we modified several times the annotation in Kubernetes, so the flow alternated between the two tunnels. We can see an example of this operation in fig. \ref{fig:tunnel_bw_switch}, that shows the bandwidth in both tunnels. We can appreciate that while one tunnel has traffic, the other doesn't, and that the throughput never exceeds 3 Mbps in the rate-limited tunnel. 

Finally, we measured the delay between changing the application annotation in Kubernetes until traffic appeared in the other tunnel. We considered that traffic appeared in the other tunnel when it received more than 1 Mbps. We repeated the experiment 200 times, and the delay was below 23 seconds 80\% of the times, with a maximum of 31 seconds.

\section{Related Work}

We can find several works discussing network autoscaling, but they are usually limited to a specific domain, such as the datacenter \cite{predictable_datacenter_networks}, virtual network functions  \cite{VNF_ML}, or video streaming \cite{videodemand}. However, in our case we focus on the WAN, and take an architectural approach instead of an algorithmic one because we make use of existing knowledge from the cloud to integrate it in the network policies.

Regarding application-aware networks, a classical approach has been extending the socket API \cite{socketIntents}. Contrarily, we do not modify the host stack but use an API-driven approach to make the network aware of application requirements. Other works in a similar direction to ours focus on the connections between end users and ISPs \cite{pavingFirstMileQoS}, or inter-datacenter connections \cite{RodriguezYaguache2019EdgeComputingContainer}, while  our proposal is centered on cloud applications and WANs. A more recent  proposal suggests merging the Layer 3 (IP) stack with the Layer 7 (application) stack, i.e., service meshes, in order to increase performance and application awareness of the network \cite{fullStackSDN}. 

With respect to labeling applications with networking requirements, DAWN \cite{dawnCoProgrammingDistributedApplication} suggests adding annotations directly in application code, in order to influence network behavior and implement complex network control policies. In addition, a different approach leverages the importance of Content Distribution Networks (CDNs) to reduce  the  distribution cost when using different CDNs, and maintain or increase client QoE at the same time \cite{costPerformanceOptimizeMultihoming}. Finally, the IETF Application Layer Traffic Optimization (ALTO) protocol has been proposed to optimize WAN 
traffic in edge computing scenarios \cite{altoTrafficOptimizationEdgeComputing}. In a nutshell, ALTO collects network information from the different edge computing clusters, that is later used to allocate edge computing workloads.





\section{Open Research Challenges}
\label{Conclusions}

The softwarization of the WAN and the advent of connectivity service providers that are exposing APIs to dynamically provision underlay capacity \cite{PacketFabricNaasBlog}, together with the development of an application-to-network API makes proactive autoscaling of WAN resources as easy as autoscaling compute and  memory. This is a very significant improvement, compared with the status quo of overprovisioning WAN capacity to address peaks of demand, as well as from solutions that monitor network utilization reacting to congestion and saturation.  The basic prototype of WAN autoscaling that we implemented shows that it is possible to extend the scope of autoscaling to the WAN, but we wish to conclude this paper with a list of future research challenges that are worth further investigation.

\textbf{Bandwidth Estimation:}  regarding vertical network autoscaling, it is crucial to perform an accurate estimation of the bandwidth required by each compute replica, in order to maximize efficiency, prevent congestion or avoid overprovisioning bandwidth, and its associated cost. While in our prototype we let the developers configure this parameter, a more accurate approach is to automatically estimate it using sophisticated techniques. There is extensive research on bandwidth estimation \cite{bwEstSurvey}, with monitoring tools or prediction algorithms, either linear or non-linear prediction models. Additional research is needed to understand how such existing techniques can benefit from having information from the cloud orchestrator.

\textbf{Bandwidth Granularity:} the bandwidth allocation granularity offered by the overlay controller is key to avoid overprovisioning. That is, the overlay controller can increase or decrease the allocated bandwidth in certain steps (e.g., 10, 50, 100 Mbps). Larger granularities that can match the required bandwidth of the compute replicas will lead to very efficient use of the resources. However, offering large granularities (e.g., steps of 1 Mbps) complicates the SDN overlay controller management. In addition, very large granularities (e.g., 1 kbps) can lead to issues handling traffic bursts. As a result, additional research is required in this field.

\textbf{Provisioning  Time:} in the context of underlay providers, the maximum capacity of a tunnel is typically a management plane configuration parameter. In many network scenarios, the provisioning time (i.e., the time until the setting takes effect) is not considered relevant and has been often done manually. In recent years programmability has allowed this time to be significantly faster, changing from manual (e.g., through phone calls) to tenths of seconds \cite{PacketFabricNaasBlog}. Despite this huge improvement, in our scenario this provisioning delay would need to commensurate with the reaction time of the cloud autoscaling engine (currently few seconds), in order to avoid under-utilization or service disruption. Additional research efforts are required to address this challenge.

\textbf{Application Diversity:} in our experiments, we have inherently assumed that each application is assigned to a single tunnel. However, in real-world scenarios, different applications can share the same connection. In order to address this, the overlay control plane should take into account that multiple applications share the same physical link, and scale capacity accordingly.

\textbf{Network Topology:} our solution is designed for scenarios with point-to-point connections, where all the end-users are connected through a single tunnel to the datacenter running the application. With this setup, autoscaling events for an application only affect one tunnel. However, in real-world scenarios, a datacenter  servicing end-users can be attached to different access networks and thus, each access network uses a different tunnel towards the datacenter. In this case, upon a surge in application traffic, and from the perspective of the cloud orchestrator, we don't know which virtual circuit needs additional bandwidth. Hence, we need more information about the traffic to determine which connections must be scaled. For example, we could use traffic monitoring techniques, or take a more proactive approach by tracking the incoming connections to the cloud datacenter.

\textbf{Billing Model:} WAN autoscaling affords a new consumption model for WAN capacity where bandwidth is provisioned (and de-provisioned) on demand in quasi real time, rather than through overprovisioning. The billing model offered by today’s underlay connectivity providers reflects the overprovisioning consumption model offering increasing SLAs (such as 1, 10, 50 Gbps) typically billed on a monthly basis. WAN autoscaling may drive the introduction of a new billing model where WAN consumption is billed on demand. We believe this may lead to a more efficient use of network infrastructure resources for the service provider, that will in turn be reflected to a lower cost for the end user. Future work in this area should include a detailed analysis of billing models for WAN autoscaling, and their comparison with current billing models.

\begin{acks}

The authors would like to thank Lorand Jakab and Elis Lulja for their help with the  deployments and the evaluation section. Thanks also to the anonymous reviewers for their feedback, and to our shepherd Mohamed Boucadair for his guiding.

This work was supported by the Spanish MINECO under contract TEC2017-90034-C2-1-R (ALLIANCE), the Catalan Institution for Research and Advanced Studies (ICREA).

\end{acks}

\bibliographystyle{ACM-Reference-Format}
\bibliography{reference}

\end{document}